\documentclass[conference]{IEEEtran}
\IEEEoverridecommandlockouts

\usepackage[numbers,sort&compress,square]{natbib}
\usepackage{amsmath,amssymb,amsfonts}
\usepackage{algorithmic}
\usepackage{graphicx}
\usepackage{textcomp}
\usepackage{xcolor}

\usepackage{tabularx}
\usepackage{enumitem}
\usepackage{balance}

\usepackage{varwidth}
\usepackage{tcolorbox}
\tcbuselibrary{breakable, skins, theorems}
\newtcbtheorem{researchquestion}{RQ}{
    varwidth boxed title,
    enhanced,
    attach boxed title to top left={xshift=2mm,yshift=-3mm}, 
    boxed title style={colframe = black, colback = white},
    coltitle = black,
    colback = white,
    colframe = black,
    fonttitle = \bfseries,
    breakable = true,
    top = 4mm
}{rq}
\usepackage{hyperref}
\newcommand{\etal}{\textit{et~al.\ }}
\newcommand{\etc}{\textit{etc.\ }}
\newcommand{\dq}[1]{``#1''}

\def\BibTeX{{\rm B\kern-.05em{\sc i\kern-.025em b}\kern-.08em
    T\kern-.1667em\lower.7ex\hbox{E}\kern-.125emX}}
\begin{document}

\title{A Method for Assisting Novices Creating Class Diagrams Based on the Instructor's Class Layout\textsuperscript{\textdagger}\\
\thanks{
  \hrule \vspace{.5\baselineskip} \noindent
  \textdagger{} This is an English translation of the paper that originally appeared in \href{https://www.jstage.jst.go.jp/browse/jssst/-char/en}{Computer
  Software}, Vol.42, No.3 (2025), pp.56--70, doi:\href{https://doi.org/10.11309/jssst.42.3_56}{10.11309/jssst.42.3\_56}.
  This is an unofficial translation by the authors of the original Japanese version.
  
  Notice for the use of this material: The copyright of this material is retained by the
  Japan Society for Software Science and Technology (JSSST). This material is published
  on this web site with the agreement of the JSSST. Please comply with Copyright Law of
  Japan if any users wish to reproduce, make derivative work, distribute or make
  available to the public any part or whole thereof.
}
}

\author{\IEEEauthorblockN{Yuta Saito}
\IEEEauthorblockA{\textit{\textit{Dept. of Information Technology and Media Design}} \\
\textit{Faculty of Advanced Engineering} \\
\textit{Nippon Institute of Technology}\\
Saitama, Japan \\
2248010@stu.nit.ac.jp}
\and
\IEEEauthorblockN{Takehiro Kokubu}
\IEEEauthorblockA{\textit{Dept. of Information Technology and Media Design} \\
\textit{Faculty of Advanced Engineering} \\
\textit{Nippon Institute of Technology}\\
Saitama, Japan
}
\and
\IEEEauthorblockN{Takafumi Tanaka}
\IEEEauthorblockA{\textit{Dept. of Software Science} \\
\textit{Faculty of Engineering} \\
\textit{Tamagawa University}\\
Tokyo, Japan \\
tanaka\_t@eng.tamagawa.ac.jp
}
\and
\IEEEauthorblockN{Atsuo Hazeyama}
\IEEEauthorblockA{\textit{Dept. of Technology and Information Science} \\
\textit{Faculty of Education} \\
\textit{Tokyo Gakugei University}\\
Tokyo, Japan \\
hazeyama@u-gakugei.ac.jp}
\and
\IEEEauthorblockN{Hiroaki Hashiura}
\IEEEauthorblockA{\textit{Dept. of Data Science} \\
\textit{Faculty of Advanced Engineering} \\
\textit{Nippon Institute of Technology}\\
Saitama, Japan \\
hashiura@nit.ac.jp}
}

\maketitle
\thispagestyle{plain}
\pagestyle{plain}

\begin{abstract}
  Nowadays, modeling exercises on software development objects are conducted in higher education institutions for information technology.
  Not only are there many defects such as missing elements in the models created by learners during the exercises,
  but the layout of elements in the class diagrams often differs significantly from the correct answers created by the instructors.
  
  In this paper, we focus on the above problem and propose a method to provide effective support to learners during modeling exercises
  by automatically converting the layout of the learner's class diagram to that of the instructor,
  in addition to indicating the correctness of the artifacts to the learners during the exercises.
  The proposed method was implemented and evaluated as a tool,
  and the results indicate that the automatic layout conversion was an effective feedback to the learners.
\end{abstract}


\section{Introduction}
In higher education institutions specializing in information technology, modeling exercises are conducted to develop learners' ability to design and construct information systems. The primary objective of these modeling exercises is to deepen learners' understanding of concepts and allow them to accumulate knowledge empirically through constructivist tasks \cite{ben}.

In modeling exercises, diagrammatic languages such as ERD, DFD, and UML (class diagrams) are commonly used as notation methods for deliverables. Chren \etal \cite{mistakes} have pointed out that UML class diagrams created by learners in modeling exercises contain significantly more errors compared to other diagrams.

The cause of such issues can be attributed to the unique characteristics of these exercises. Even in software development conducted outside educational institutions, modeling remains an essential task. In such cases, requirements, which serve as the starting point for modeling, are provided through various means, including documentation, verbal communication, and implicit methods, by both project owners and clients. Additionally, if developers are uncertain about whether they have accurately captured the requirements, they can usually ask clarification questions to the requirement providers or conduct interviews to gather additional information as needed.

On the other hand, in the modeling exercises targeted in this study, learners often receive only a brief requirement document from the instructor at the beginning of the exercise. Furthermore, they must conduct modeling exercises under the following constraints:
\begin{itemize}
  \item The instructor has prepared a correct model that aligns with the requirement document.
  \item Identifying classes not explicitly stated in the requirement document is not the focus.
  \item Determining an appropriate class arrangement independently is not the focus.
  \item The exercises are conducted within lectures at higher education institutions.
  \item The ratio of instructors to learners is high.
  \item Other models are not created simultaneously.
\end{itemize}
As a result, it is challenging for learners who have taken lectures on modeling but have no prior experience with modeling exercises (hereafter referred to as \dq{beginners}) to create models under these constraints.

Additionally, there are issues with the feedback method provided to beginners at the end of the modeling exercises. Many instructors present a correct class diagram as an example to learners at the end of the exercise and use it as a reference for explaining the problems. However, since these explanations are delivered unilaterally from the instructor to the learners, learners may not fully grasp the intent behind the presented correct model or explanations. Consequently, when learners attempt to tackle new exercises, their previous lack of understanding prevents them from accurately interpreting the intent of the exercise problems. This leads to the recurring issue of creating deliverables with missing or incorrect elements \cite{mistakes}.

The challenges beginners face in modeling, including those mentioned earlier, can be attributed to the following issues \cite{ogiso,aStudy}:
\begin{itemize}
  \item Lack of knowledge regarding UML terminology and design methods
  \item Insufficient understanding of how to use modeling tools
  \item Inadequate comprehension of the system domain and specifications described in the problem statement
  \item Insufficient feedback from instructors, mainly due to time and effort constraints
  \item Insufficient feedback from the development environment, such as output and error messages, compared to programming exercises
\end{itemize}

This study focuses on the issue that, compared to programming exercises, modeling exercises provide insufficient feedback from the development environment. However, it has also been pointed out that providing excessive feedback can be equivalent to presenting the correct solution, potentially hindering the learning effectiveness of students \cite{using}. 

As an approach to addressing these issues simultaneously, Ichinohe \etal\cite{ichinohe} proposed an environment where learners receive a numerical similarity score ranging from 0.0 to 1.0, indicating how closely their work resembles the correct model. This allows learners to refine their models through trial and error without relying directly on a correct example.

However, feedback based solely on similarity scores can sometimes lead to a disconnect between learners' perceptions and their actual results. Since class diagrams used in modeling exercises do not contain an overwhelming number of elements, consider the following example: if a diagram class has three attributes, with one attribute correctly matching the correct model and two attributes being incorrect, the similarity score displayed to the learner would be 0.33. 

If a learner has correctly identified one of the three attributes from an extensive solution space, they should ideally use it as a clue to infer the remaining incorrect attributes and improve their similarity score. However, learners may instead feel that their work has been entirely rejected, leading them to abandon their current model and attempt to reconstruct it from scratch. As a result, their similarity score may fluctuate irregularly instead of showing steady improvement.

In this study, to further address the aforementioned issues, we focused on the layout of models. 

In the field of reverse engineering, numerous studies have examined the impact of class diagram layouts on readers' understanding. For example, Störrle \cite{layout_ex} pointed out that following layout rules and aesthetic principles improves the readability of class diagrams. 

Layout rules range from fundamental principles, such as avoiding overlapping elements in class diagrams, to more specific rules, such as placing parent classes at the center of child classes \cite{oya_ko}. Additionally, in design patterns, rules have been proposed for grouping related classes together (clustering), among other class diagram-specific guidelines.

In this paper, we propose a method to support learners' modeling by utilizing class diagram layouts. 

This method uses the layout of class diagrams (correct examples) as a means of support to address the challenges learners face, as mentioned above. In this paper, the term \dq{layout} refers to the arrangement of classes.

The objective of this study is to investigate the following two aspects:
\begin{enumerate}
  \item The impact of layout on the similarity of learners' deliverables
  \item The transition of similarity scores during the modeling exercise
\end{enumerate}
By examining these aspects, this study aims to determine whether using class diagram layouts as support can provide beneficial feedback in modeling exercises.

The structure of this paper is as follows. 
Section 2 introduces studies related to class diagram creation support and class diagram layouts.
Section 3 explains the proposed method, which involves automatic layout transformation.
Section 4 describes the research questions (RQs) and the evaluation methods.
Section 5 presents the experimental results and provide discussions.
Section 6 discusses potential threats to the validity of this study.
Section 7 concludes the paper and discusses future work.

\section{Related Work}
\subsection{Support for Class Diagram Creation}
Numerous studies have been conducted on automatically providing feedback to learners on their deliverables in modeling exercises.
Examples are shown in Table \ref{tab:paper_list}.
\begin{table}[tb]
  \centering
  \normalsize
  \caption{Classification of Studies on Supporting Class Diagram Creation}
  \begin{tabular}{l|p{3cm}|p{3cm}}
    \hline
    \# & \textbf{Classification} & \textbf{References} \\
    \hline
    1 & Detection of Missing Elements & \cite{daria,Hasker,stephan,tobias,Schots,web,wei} \\
    \hline
    2 & Automatic Scoring & \cite{auto,younes,ichinohe,miyashima,tool} \\
    \hline
    3 & Others & \cite{nemo,using,ar,3d} \\
    \hline
  \end{tabular}
\label{tab:paper_list}
\end{table}

Research by Wei \etal\cite{wei} is cited as a study that detects missing elements.
Wei \etal\cite{wei} proposed a method and framework to discover missing requirements (semantic holes) from class diagrams, sequence diagrams, and activity diagrams created by the same learner.
This method has the advantage of improving the consistency and completeness of the learner's created models without using correct examples.
On the other hand, since this method focuses on inconsistencies in notation that occur when learners create multiple deliverables, it cannot detect requirements that the learner completely overlooks and are not described in any of the models as semantic holes.
Therefore, it is not suitable for supporting only class diagram modeling exercises, which is what this research assumes, or for complete beginners.

Soler \etal\cite{web} developed a web-based tool to support the teaching and learning of UML class diagrams.
This tool gives learners practice problems and automatically corrects their deliverables.
After correction, learners can receive feedback that summarizes the detected differences.

Krusche \etal\cite{stephan} created an online modeling editor.
Krusche \etal's \cite{stephan} editor can highlight missing requirements next to elements, allowing learners to receive individual and direct feedback.

A common advantage of these missing element detection methods is that by explicitly pointing out missing elements, learners can correct their deliverables themselves, 
thereby improving their motivation to learn. On the other hand, there are issues such as learners not being able to receive feedback during modeling or not being able to receive feedback multiple times.

Numerous studies have been conducted on automatically scoring deliverables, particularly class diagrams \cite{auto,younes,ichinohe,miyashima,tool}.
Ichinohe \etal\cite{ichinohe} developed a tool that returns the similarity between a correct example and the deliverable as feedback to the learner.
This has made it possible for learners to receive feedback during modeling.
However, as mentioned in the previous section, if the feedback deviates from the learner's actual understanding, it is difficult to consistently improve the similarity.

Karasneh \etal\cite{using} proposed a method to utilize a case study collection for software design in modeling exercises, stating that the case study collection (UML repository) helps improve the quality of the learner's models.
On the other hand, they point out that with such methods, there are learners who cannot distinguish between good and bad examples, and that learners may feel that using examples is \dq{the instructor's negligence} or that \dq{examples are more than necessary to learn the current content.}

In addition to these, Rodrigues \etal\cite{3d} and Reuter \etal\cite{ar} propose to intuitively understand elements by representing models (class diagrams) in AR or 3D.
Furthermore, Rocco \etal\cite{nemo} propose a system that uses neural networks to suggest the next modeling action.

While various studies have been conducted to support learners in creating class diagrams in this way, there are few studies that allow for receiving feedback multiple times during modeling and that confirm that the feedback is at an appropriate level as a hint.

\subsection{Class Diagram Layout}

As mentioned earlier, in the field of reverse engineering, there has been discussion about what kind of layout is preferable when automatically generating class diagrams \cite{layout_ex,Adrian,Bonita}.

Störrle \cite{layout_ex} clarified that a \dq{good} layout that adheres to layout standards has beneficial effects on several aspects of cognitive load. In particular, they point out that novice modelers benefit more from the layout than advanced modelers.

Adrian \etal\cite{Adrian} proposed a consistent layout for software visualization. In this layout, the position of software artifacts reflects their vocabulary, and the distance corresponds to the similarity of the vocabulary. By placing the vocabulary of entities in an m-dimensional space and using Multidimensional Scaling to map these positions to a 2D display, a software map with digital elevation and contour lines is ultimately created.

Bonita \etal\cite{Bonita} clarified that clustering design patterns improves the accuracy of role detection. They also showed that there are differences in the eye movements of beginners and experienced users.

Helen \etal\cite{helen} state that if the notation does not match the creator's intuition, there is a lack of confidence and they will look at the class diagram more intently, making it easier for subjects to identify errors.

Ahmar \etal\cite{Ahmar} propose a method to change layers according to the situation to improve readability. By using these layers, it is possible to add information such as background color, brightness, line thickness, and text font to the elements of the original class diagram. There is also a layer that makes the original classes invisible, reducing the number of displayed classes. By changing the layers according to the situation, the complexity of large class diagrams is reduced, and communication becomes more effective.

As such, it is clear that the quality of class diagram layout affects understanding. On the other hand, although these layout rules are utilized in fields such as automatic class diagram generation and reverse engineering, there is little research that applies them to the modeling field.

\subsection{Similarity by Tanaka \etal}
In this paper, we use the similarity by Tanaka \etal\cite{similarity} to calculate the similarity between the deliverables and the correct examples in the proposed method, experiments, and evaluation. The following \textit{snapshot} of Tanaka \etal corresponds to the deliverables in this research.

\begin{tcolorbox}[enhanced jigsaw,
  colframe = black!40,
  colback = white,
  breakable = true,
  parbox = false]

Below, we denote the \textit{snapshot} as $S$ and the correct example class diagram as $A$.
Also, we use $nameSim(E_1, E_2)$ (a function that returns the similarity of element names of elements $E_1, E_2$ in the range of 0 (minimum) to 1 (maximum)) as a function to calculate the similarity of element names (class name, attribute name, relationship name) of class diagram elements (class, attribute, relationship).
Currently, we are considering using $nameSimilarity$ defined in Xing \etal\cite{Xing2005} as $nameSim$.
$nameSimilarity$ is a function that calculates the number of \dq{adjacent 2 characters} common to the strings being compared and returns the value obtained by dividing twice that number by the sum of the string lengths of the strings being compared. It is stated that this is more suitable for comparing element names than the conventionally used $LCS$ ($Longest$ $Common$ $Subsequence$) algorithm.
\begin{description}
  \item[(1) Comparison of classes]
\end{description}
\par
\noindent\textcircled{\scriptsize 1} Extraction of comparison target classes and calculation of class name similarity $CNS$ ($Class$ $Name$ $Similarity$)

Let $C_i \in S$ and $C_j \in A$. A pair of $C_i$ and its corresponding $C_j$ (the target for similarity calculation) is called a comparison target class pair. The calculation procedure is as follows:

\begin{enumerate}[label=\Roman*]
  \item For a certain $C_i$, calculate the class name similarity \texttt{$nameSim$}($C_i$,$C_j$) with all $C_j$ that do not belong to any comparison target class pair.
  \item If the maximum value of the calculated class name similarity $\texttt{$max$}(\texttt{$nameSim$}(C_i,C_j))$ exceeds the threshold $T_n$, then that $(C_i,C_j)$ is extracted as a comparison target class pair. Also, the class name similarity $CNS$ of $(C_i,C_j)$ is set to \texttt{$nameSim$}($C_i$,$C_j$).
  \item Perform steps I and II for all $C_i$.
  \item Consider any $C_j$ that does not belong to a comparison target class pair as a missing class, and calculate the number of missing classes as $NMC$ ($Number$ $of$ $Missing$ $Classes$).
\end{enumerate}

\noindent\textcircled{\scriptsize 2} Calculation of attribute name similarity $ANS$ ($Attribute$ $Name$ $Similarity$)

For each attribute of the comparison target class pair $(C_i, C_j)$, extract the comparison target attribute pair (the pair of attributes that are the target for similarity calculation) using the same procedure as for classes, and calculate the attribute name similarity ANS for each pair. Also, the attributes of $C_j$ that do not belong to the comparison target attribute pair are considered missing attributes, and their number is calculated as $NMA$ ($Number$ $of$ $Missing$ $Attributes$).

\noindent\textcircled{\scriptsize 3} Calculation of class similarity $CS_{i}$ ($Class$ $Similarity$)

The similarity $CS_{i}$ between a certain class $C_i \in S$ and the correct example is calculated by the following formula if class $C_i$ belongs to a comparison target class pair, and 0 otherwise. In the formula, $NA$ represents the number of attributes of $C_i$.

\begin{equation}
  CS_{i} = \frac{CNS + \sum ANS}{1 + NA + NMA} \notag
\end{equation}

The similarity $CS_{all}$ for all classes in $A$ and $S$ is calculated by the following formula. In the formula, $NC_{S}$ represents the number of classes in $S$.

\begin{equation}
  CS_{all} = \frac{\sum CS_{i}}{NC_{s} + NMC} \notag
\end{equation}

\begin{description}
  \item[(2) Comparison of relationships]
\end{description}

\noindent\textcircled{\scriptsize 1} Extraction of comparison target relationships
 
From relationship $R_i \in S$ and relationship $R_j \in A$, extract $(R_i, R_j)$ as a pair of comparison target relationships where the pair of classes at their respective ends is a pair of comparison target classes. Associations $R_j$ that are not included in the pair of comparison target relationships are considered missing relationships, and the number of missing relationships is calculated as $NMR$ ($Number$ $of$ $Missing$ $Relationships$).
 
\noindent\textcircled{\scriptsize 2} Calculation of relationship name similarity $RNS$ ($Relationship$ $Name$ $Similarity$)

For a pair of relationships belonging to the comparison target relationships, if \texttt{$nameSim$}($R_i$,$R_j$) exceeds the threshold $T_n$, the relationship name similarity RNS of $(R_i,R_j)$ is \texttt{$nameSim$}($R_i$,$R_j$), otherwise it is 0.

\noindent\textcircled{\scriptsize 3} Calculation of multiplicity similarity $CaS$

For the corresponding ends of the pair of relationships $(R_i,R_j)$ belonging to the comparison target relationships, the multiplicity similarity $CaS$ ($Cardinality$ $Similarity$) is determined. In the class diagrams targeted this time, we will use four types of multiplicities: \dq{1,} \dq{0..1,} \dq{1..*,} and \dq{*,} and set the similarity for these combinations in advance. Determine the corresponding similarity value according to the pair of multiplicities being compared.

\noindent\textcircled{\scriptsize 4} Calculation of relationship similarity $RS_{i}$ ($Relationship$ $Similarity$)

The similarity $RS_{i}$ between relationship $R_i \in S$ and the correct example is calculated by the following formula if the target relationship $R_i$ belongs to the comparison target relationships, and 0 otherwise.

\begin{equation}
 RS_{i} = \frac{RNS + \sum CaS}{1 + 2} \notag
\end{equation}
 
The similarity $RS_{all}$ for all relationships in $A$ and $S$ is calculated by the following formula. In the formula, $NR_{S}$ represents the number of relationships in $S$.

\begin{equation}
 RS_{all} = \frac{\sum RS_{i}}{NR_{s} + NMR} \notag
\end{equation}

\begin{description}
 \item[\parbox{6cm}{(3) Calculation of class diagram similarity $CDS$ ($Class$ $Diagram$ $Similarity$)}]
\end{description}

The similarity $CDS$ of the class diagram pair $(S,A)$ is calculated by the following formula.
\begin{equation}
 CDS = \frac{CS_{all} + RS_{all}}{2} \notag
\end{equation}
 
\end{tcolorbox}

\section{Proposed Method}

\begin{figure*}[tb]
  \begin{center}
    \includegraphics[width=0.95\textwidth]{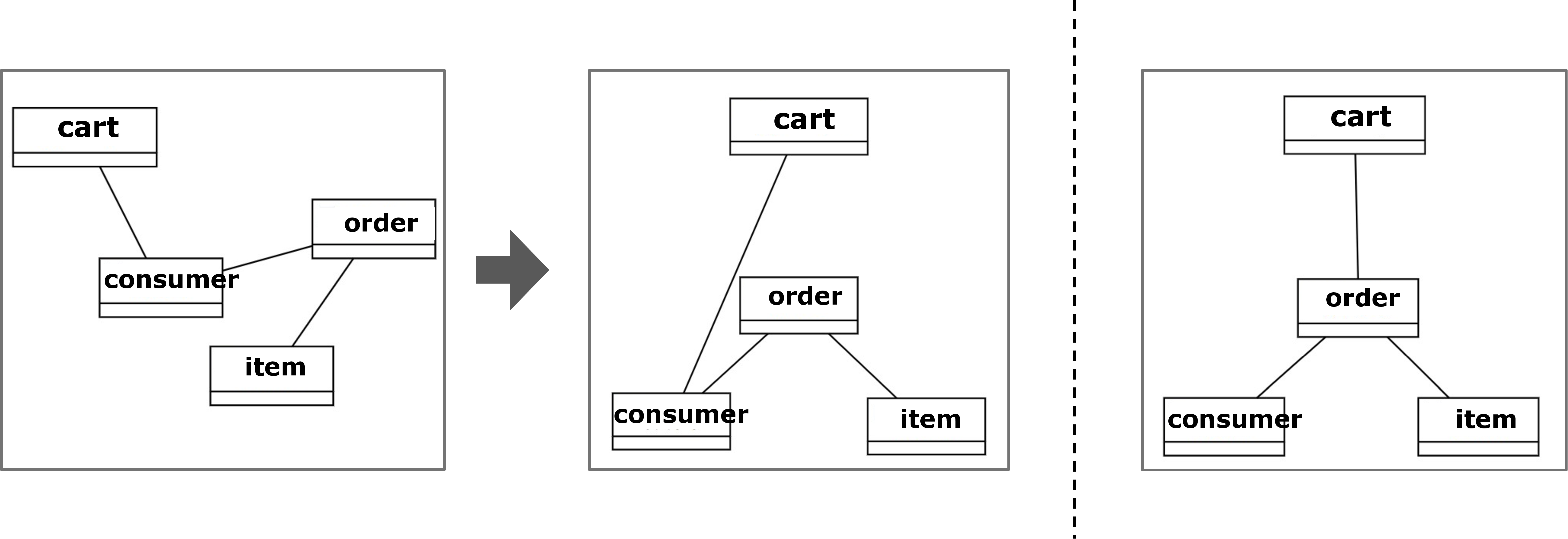}
    \caption{Automatic layout conversion (left: student's work, center: student's work after automatic layout conversion, right: correct example)}
    \label{relocation}
  \end{center}
\end{figure*}

In this paper, we automatically convert the layout of the student's work to the layout of the correct example.
Specifically, it involves rearranging the placement of classes in the class diagram (student's work) created by the learner so that it becomes the same as the placement of classes in the correct example.
This section explains the automatic layout conversion.
An example of automatic layout conversion is shown in Figure \ref{relocation}.
The right side is the correct example created by the instructor, the left side is the student's work,
and the center is the student's work with its layout changed to match the layout of the correct example (right).
In the automatic layout conversion, the classes in the student's work (Cart, Customer, Order, Product) are rearranged to the coordinates of the classes in the correct example (Cart, Customer, Order, Product).
At this time, classes and relationships are not created; only the classes present in the student's work are used for rearrangement.
Therefore, the student's work after automatic layout conversion may have missing or extra classes, attributes, and relationships.
The learner is encouraged to correct errors in their current work by re-examining their own work from the perspective of the instructor who created the correct example.

The automatic layout conversion can be performed instantaneously, and learners can perform it as many times as they want during the exercise.
The only information feedback to the learner through automatic layout conversion is the layout of the correct example; they cannot directly obtain information such as the similarity calculated during the layout conversion process.

The method of changing the layout is broadly divided into three steps.
These are \dq{searching for corresponding classes,} \dq{rearranging classes,} and \dq{restoring relationships while maintaining the original relationships.}
Each of these will be described below.

\subsection{Searching for Corresponding Classes}
First, we will explain \dq{searching for corresponding classes.}
This is because, in order to perform automatic rearrangement, it is necessary to identify how each element included in the correct example and the student's work corresponds.

In the search for corresponding classes, we use the class-by-class similarity $CS$ to find \dq{corresponding classes} from the classes in the student's work.
The definition of a corresponding class is that the value of $CS$ for a certain class in the correct example is 0.4 or more and is the largest among the classes in the student's work.
If there are multiple classes with the maximum $CS$ value, the one with the lexicographically earlier class name is given higher priority.
In other words, for one class in the correct example, only one corresponding class is selected from the classes in the student's work.

\subsection{Rearranging Classes}
Next, we will explain \dq{rearranging classes.}
In the coordinate change, only the coordinate data of the classes in the student's work is rewritten in order to rearrange the corresponding classes found in the first step from the student's work to the coordinates of the corresponding classes in the correct example.
Then, for the classes in the student's work that were not determined to be corresponding classes in the first step (non-corresponding classes), they are rearranged to the top-left of the drawing area, at coordinates $(0,0)$.

The reason for performing such a rearrangement for non-corresponding classes is to avoid the following two problems. The first is that if non-corresponding classes are left in their original positions, they may overlap with the rearranged classes, potentially causing learners to confuse non-corresponding classes with corresponding classes.
The second is that if learners continue modeling work from such a state, the rearranged corresponding classes may be moved again by the learners.
Note that if multiple non-corresponding classes exist, they will be displayed overlapping each other.

\subsection{Restoring Associations While Maintaining Original Relationships}
Finally, we will explain \dq{restoring relationships while maintaining original relationships.}
In the automatic layout conversion, since only the coordinates of the classes are changed, the relationships connecting the classes are restored so as not to disrupt the original relationships.
This is to prevent learners from becoming unable to recognize the class diagram as one they created themselves due to the automatic layout conversion.
At this time, relationships with non-corresponding classes that have been rearranged to coordinates $(0,0)$ also remain, so there is a possibility that relationships and classes will overlap.

The automatic layout conversion is completed by going through these three steps.

\subsection{Implementation}\label{implement}
In this paper, we extended the model drawing tool \textit{KIfU}\cite{KIfU} to add an automatic layout conversion function.
\textit{KIfU} already has a function to calculate the similarity of class diagrams.
We decided to implement the function by adding a feature to associate the classes in the correct example with the corresponding classes in the student's work.

In \textit{KIfU}, every time a learner edits a class diagram, the information of the class diagram is registered in the database as a snapshot.
In this research, by extending this snapshot to change the coordinate values of the classes, we made it possible to convert the layout so that the learner's classes correspond to the correct example.
The relationships included in the aforementioned snapshot do not have coordinates and only hold information about the classes at both ends, so they are not affected by the process of changing the coordinates of the classes mentioned above.
Therefore, the original relationships can be maintained without specifically extending the functionality.
Furthermore, to allow learners to perform automatic layout conversion at any time, we made it so that a series of steps is automatically executed by button operation.

Since similarity is important for performing automatic layout conversion, we added a color-based feedback function for class names and attribute names to prevent a decrease in similarity due to typing mistakes or synonyms.
The color of each element name changes to red if the class name or attribute name completely matches the correct example ($nameSim=1.0$), to black if it partially matches ($0.0<nameSim<1.0$), and to blue if it completely mismatches ($nameSim=0.0$).
This allows learners to know to what extent their work matches the correct example.

\subsection{Feedback Obtained by Tool Users}
Learners using \textit{KIfU}\cite{KIfU}, which implements the proposed method, can obtain feedback from the following visual effects by pressing a check button.

\begin{enumerate}[label=\Alph*)]
 \item Classes rearranged to the coordinates of the correct example
 \item Classes rearranged to coordinates $(0,0)$
 \item Unnatural whitespace
 \item Associations that intersect with other elements
 \item Color of element names (red, black, blue)
\end{enumerate}

A) to D) are provided by the automatic layout conversion function.
From A), \dq{Classes rearranged to the coordinates of the correct example,} learners can recognize that this class is included in the correct example.
From B), \dq{Classes rearranged to coordinates $(0,0)$,} learners can recognize that this class is not included in the correct example.
From C), \dq{Unnatural whitespace,} learners can indirectly recognize the possibility that missing elements (classes, relationships) exist there.
From D), \dq{Associations that intersect with other elements,} learners can recognize that intersected relationships are inappropriate relationships.

\begin{figure}[tb]
  \begin{center}
    \includegraphics[width=0.45\textwidth]{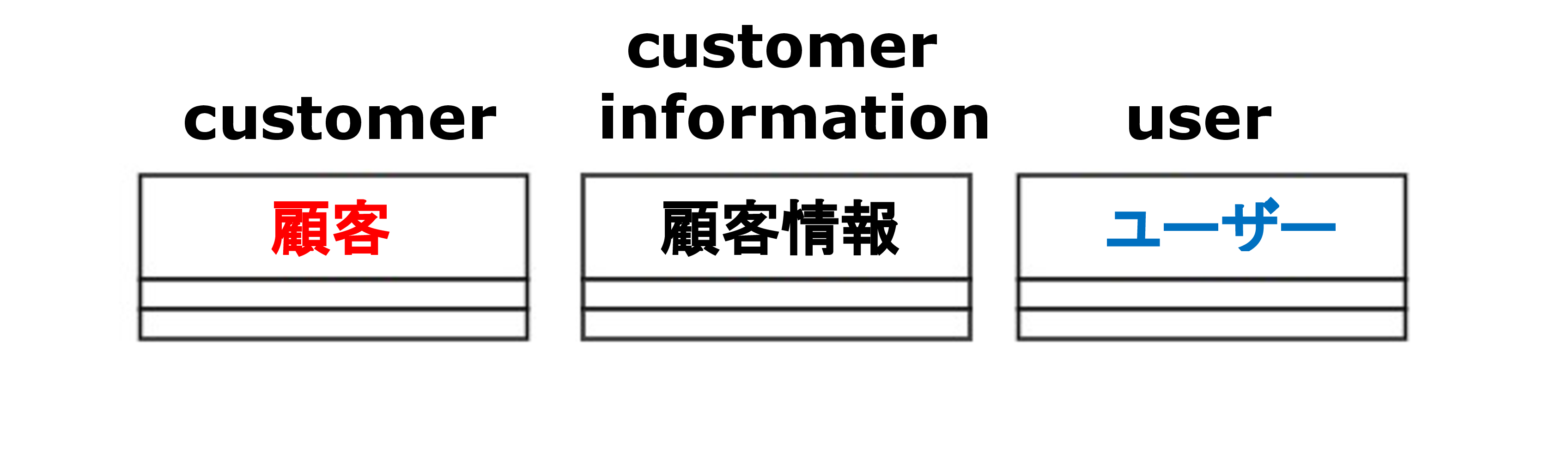}
    \caption{Specific example of feedback using the color of element names}
    \label{color_feedback}
  \end{center}
\end{figure}

And E) is provided by the color-based feedback function mentioned in Section \ref{implement}.
Figure \ref{color_feedback} shows a specific example of feedback using the color of element names.
This allows learners to recognize whether elements they created are the same as or similar to elements that exist in the correct example, based on the color of the element names (red, black, blue).
For example, even if the class placement is exactly the same as the correct example, if the learner answers \dq{Customer Information} for an element whose class name in the correct example is \dq{Customer,} the class will not be rearranged and will be displayed in black, indicating that the element name partially matches.
Through this feedback, learners can effectively understand the accuracy and areas for improvement of their own work.

\section{Evaluation}

\subsection{Research Questions (RQs)}
We will evaluate the proposed method by answering the following research questions.

\begin{researchquestion}{}{rq1}
 How does the layout of the correct example effect the creation of the student's work?
\end{researchquestion}
We will investigate the impact on learners' class diagrams using Tanaka \etal's similarity $CDS$. In addition to the similarity $CDS$, we will also evaluate whether the effect is on classes or relationships based on the class similarity $CS_{all}$ and the relationship similarity $RS_{all}$.    

\begin{researchquestion}{}{rq2}
 When does the influence of the layout become apparent?
\end{researchquestion}
In this RQ, we will use the similarities $CDS$, $CS_{all}$, and $RS_{all}$ as in RQ\ref{rq:rq1}. By measuring the similarity during modeling, we will evaluate whether the learners' stuck state has been resolved.

\subsection{Evaluation Methods}
We conducted an experiment to demonstrate the usefulness of automatic layout conversion for learners' class diagram modeling.
The subjects were 20 students from the department of information technology and media design at NIT who had completed a course on object-oriented programming.
All subjects were students with little to no experience in creating class diagrams.
These 20 subjects were randomly assigned to the following two groups of 10 students each.

\begin{enumerate}
 \item A group using \textit{KIfU} with the automatic layout conversion function (hereinafter referred to as the experimental group)
 \item A group using \textit{KIfU} without the automatic layout conversion function (hereinafter referred to as the control group)
\end{enumerate}

 In the experiment, subjects performed a modeling exercise in which they created a class diagram based on a system overview.
 This system overview aims to build an e-commerce system for Wakaba Trading Company to digitize product orders that were previously placed via fax or phone.
 The overview includes Wakaba Trading Company's business flow and digitalization requests.
 This overview is a modified version by the authors based on the document \cite{answer} (p.~63).

 The experimental procedure was as follows with both the experimental and control groups using the same procedure.

\begin{enumerate}
  \item Explain class diagrams and the drawing tool to the subjects.
  \item Subjects freely use the tool as a tutorial.
  \item Distribute a sheet of paper describing the system overview of the e-commerce system to the subjects.
  \item Subjects perform a modeling exercise using the tool (1 hour).
  \item Subjects submit their class diagrams within the tool.
\end{enumerate}
   
In the explanation of the drawing tool for the experimental group, in addition to explaining basic operations, we also explained the automatic conversion to the layout of the correct example.
At this time, detailed explanations such as the definition of similarity were not given, but the subjects experienced how many classes needed to match to be judged as corresponding classes by actually using the function in the tutorial.

The evaluation targets are the class diagrams created by the 20 subjects.
From these class diagrams, we calculated the similarities $CDS$, $CS_{all}$, and $RS_{all}$, and performed a two-tailed t-test to examine significant differences.
In addition, when classifying classes, since the $CS$ value is affected even if a single character in a word is different, we imposed a restriction that the words that can be used in the class diagram are limited to the words described in the problem statement.

\section{Discussion}
\subsection{RQ\ref{rq:rq1}: How does the layout of the correct example affect the creation of the student's work?}
\begin{figure}[tb]
  \begin{center}
    \includegraphics[width=0.45\textwidth]{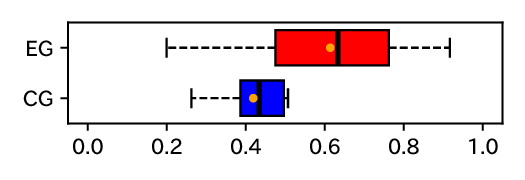}
    \caption{Comparison of $CDS$ between the experimental groups (EG) and control groups (CG)}
    \label{box_cds}
  \end{center}
\end{figure}

Figure \ref{box_cds} shows the similarity $CDS$ obtained in the modeling exercise.
As a result of the t-test, a significant difference was observed in the similarity $CDS$ ($p=0.0235,\alpha=0.05$).
This result indicates that automatic layout conversion is beneficial feedback for learners' class diagram modeling.
   
\begin{figure}[tb]
  \begin{center}
    \includegraphics[width=0.45\textwidth]{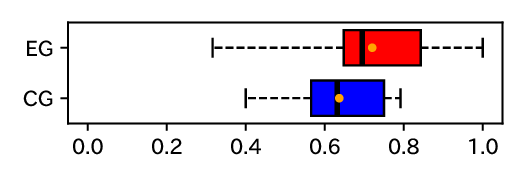}
    \caption{Comparison of $CS_{all}$ between the experimental groups (EG) and control groups (CG)}
    \label{box_cs}
  \end{center}
\end{figure}

Next, we investigated what kind of effect this has on each element of the class diagram.
Figure \ref{box_cs} shows the results of $CS_{all}$.
As a result of the test for $CS_{all}$, no significant difference was observed ($p=0.2886,\alpha=0.05$).
   
The results regarding classes were different from the authors' expectations.
The authors thought that learners would notice the existence of missing classes from the whitespace in the layout after automatic conversion, but no significant difference was observed between the experimental and control groups.
   
Next, Table \ref{cs_list} shows the results of $CS$ for each class.
Focusing on the $CS$ of each class, a significant difference was observed in the \dq{Order Item} class ($p=0.0003734,\alpha=0.05$).
Although not a significant difference, the $CS$ of the \dq{Order} class is larger in the control group than in the experimental group.
We believe that this result is due to the similarity in the class names \dq{Order} and \dq{Order Item,} which caused many learners to perceive these two concepts as one.
Automatic layout conversion can suggest the existence of missing classes from the distance between classes, but its effect on understanding concepts is considered limited.
   
\begin{table}[tb]
  \centering
  \normalsize
  \caption{Average $CS$ of each class}
  \begin{tabular}{l|l|>{\raggedleft\arraybackslash}p{2.7cm}|>{\raggedleft\arraybackslash}p{2.7cm}}
      \hline
      \# & \textbf{Class Name} & \textbf{Experimental Group} & \textbf{Control Group} \\
      \hline
      1 & Customer & 0.96 & 0.99 \\
      \hline
      2 & \textbf{Order} & 0.52 & 0.65 \\
      \hline
      3 & Cart & 0.60 & 0.48 \\
      \hline
      4 & \textbf{Order Item}\rlap{*} & 0.53 & 0.05 \\
      \hline
      5 & Product & 0.90 & 0.94 \\
      \hline
      6 & Inventory & 0.93 & 0.82\\
      \hline
      \multicolumn{4}{l}{\footnotesize *$p<0.05$}
  \end{tabular}
  \label{cs_list}
\end{table}
   
\begin{figure}[tb]
  \begin{center}
    \includegraphics[width=0.45\textwidth]{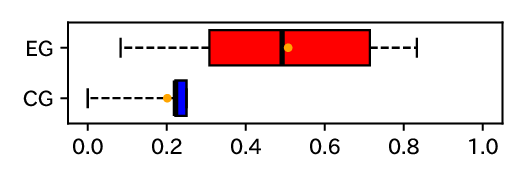}
    \caption{Comparison of $RS_{all}$ between the experimental groups (EG) and control groups (CG)}
    \label{box_rs}
  \end{center}
\end{figure}

Finally, Figure \ref{box_rs} shows the results of $RS_{all}$.
As a result of the test for $RS_{all}$, a significant difference was observed ($p=0.0050,\alpha=0.05$).
We believe this result is closely related to the presence or absence of the \dq{Order Item} class and the layout after automatic conversion.
The existence of the \dq{Order Item} class creates the possibility of creating three relationships.
We believe that the layout after automatic conversion narrows down the options for relationships that learners create.
In all the class diagrams submitted in this exercise, no overlapping elements were observed.
This indicates that learners naturally avoid overlapping elements without being explicitly told about layout rules.
From such learner behavior, we can consider that the layout is narrowing down the options.

\begin{figure}[tb]
 \includegraphics[width=0.45\textwidth]{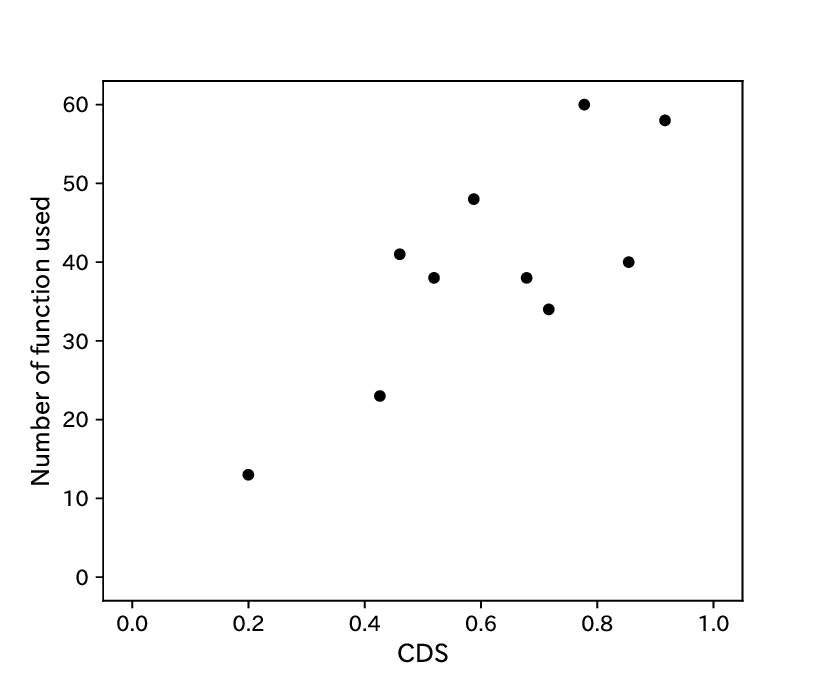}
 \caption{Distribution of the number of times the automatic conversion function was used and $CDS$}
 \label{soukan}
\end{figure}

The variance of these similarities, $CDS$, $CS_{all}$, and $RS_{all}$, is larger in the experimental group than in the control group.
To investigate the cause of this, we examined the number of times the automatic layout conversion function was used.
Figure \ref{soukan} shows the number of times the automatic conversion function was used and the distribution of $CDS$.
Correlation analysis showed a strong correlation of $r=0.7877$.
From this, it can be considered that learners who created class diagrams with higher similarity sought more feedback.

Regarding these experimental results, similar results are expected even if the instructor directly gives feedback including layout conversion verbally to the learners without making the proposed method into a tool.
On the other hand, collecting the subjects' class diagrams, performing layout conversion, and returning them requires about 1 to 3 minutes of effort per time.
In the case of this experiment, as shown in Table \ref{relocation_count}, an average of 35.5 feedbacks were given to each learner, and the instructor would need to spend at least one hour of effort per learner.
In addition, when the instructor provides feedback verbally, the problem arises that it is difficult to provide feedback in real time in modeling exercises with a large number of students, which is one of the advantages of this method.
By implementing the method as a system in this research, we were able to confirm that such problems can also be avoided.

\begin{table}[tb]
   \centering
   \normalsize
   \caption{Comparison of the number of feedbacks during the experiment}
   \begin{tabular}{c|r|r}
    \hline
    \# & \multicolumn{1}{c|}{Experimental Group} & \multicolumn{1}{c}{Control Group} \\
    \hline
    1 & 34 & 41\\
    \hline
    2 & 38 & 20 \\
    \hline
    3 & 13 & 25 \\
    \hline
    4 & 23 & 39 \\
    \hline
    5 & 60 & 45 \\
    \hline
    6 & 40 & 26 \\
    \hline
    7 & 48 & 46 \\
    \hline
    8 & 38 & 26 \\
    \hline
    9 & 41 & 42 \\
    \hline
    10 & 58 & 7 \\
    \hline
    Average & 39.3 & 31.7 \\
    \hline
    Overall Average & \multicolumn{2}{c}{35.5} \\
    \hline
   \end{tabular}
   \label{relocation_count}
\end{table}
  
Next, to confirm the learning effect of the proposed method, we conducted interviews with the subjects.
In the interviews, we selected six items from the learning items in conceptual data modeling by Kato and Nanba\cite{kato} to understand the learning content.
Table \ref{interview} shows the specific interview items and results.

\begin{table}[tb]
  \caption{List of interview items}
  \begin{center}
  \begin{tabular}{l|>{\raggedright\arraybackslash}p{4.7cm}|r}
    \hline
    \# & \textbf{Question Item} & \textbf{Positive Response} \\
    \hline
    1 & Did you understand the notation for creating class diagrams? & 7 \\
    \hline
    2 & Did you acquire the ability to extract classes? & 6 \\
    \hline
    3 & Did you acquire the ability to extract attributes? & 6 \\
    \hline
    4 & Did you acquire the ability to discover relationships between elements? & 3 \\
    \hline
    5 & Did you acquire the ability to review the class diagrams you created? & 5 \\
    \hline
    6 & Were you able to understand the real-world structure from the class diagrams you created? & 6 \\
    \hline
  \end{tabular}
  \label{interview}
  \end{center}
\end{table}
    
From \#1, \#2, \#3, \#5, and \#6 in Table \ref{interview}, it was confirmed that many students felt certain degree of learning effect.
On the other hand, in the item regarding relationship creation (\#4 in Table \ref{interview}), only a small number of students felt a learning effect.
From this, it is thought that there is an overall learning effect in the exercises using this method, but it was suggested that the learning effect in relationship creation might be weak.

Among the subjects who answered that they had gained a learning effect in relationship creation, there were opinions such as \dq{I drew relationships by thinking for myself while the instructor's layout limited the options for drawing relationships} and \dq{I became able to write relationships myself without relying on layout hints,} revealing that the support by automatic layout conversion had a certain effect. On the other hand, among the subjects who did not feel a learning effect, there were opinions such as \dq{I didn't understand why I couldn't draw relationships} and \dq{I somehow managed to draw them.}

\subsection{RQ\ref{rq:rq2}: When does the influence of the layout become apparent?}
\begin{figure}[tb]
  \begin{center}
    \includegraphics[width=0.45\textwidth]{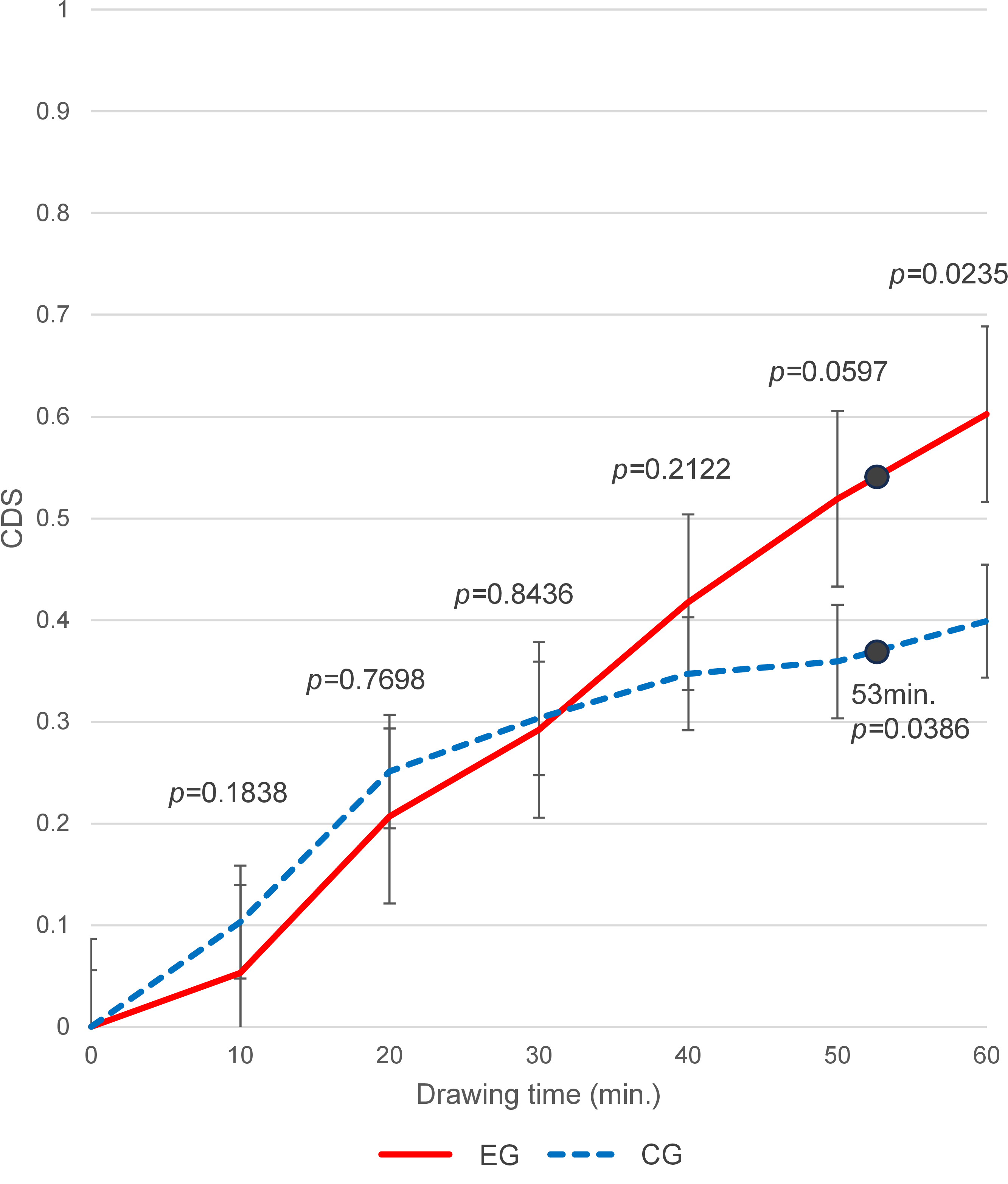}
    \caption{Transition of $CDS$ for the experimental groups (EG) and control groups (CG)}
    \label{line_cds}
  \end{center}
\end{figure}

We investigated the transition of similarity during the modeling exercise using \textit{KIfU}'s function to record the class diagram each time it is edited.
Figure \ref{line_cds} shows the change in similarity during the modeling exercise.
  
As can be seen from Figure \ref{line_cds}, there is no significant difference between the experimental and control groups until 53 minutes of drawing time.
However, after 20 minutes, when many subjects start creating relationships, the $CDS$ of the control group slows down but still continues to rise gradually, while the experimental group continues to improve similarity without slowing down.
From this result, we believe that this method is helping to resolve the learners' stuck state in modeling.
The results of the control group suggest that if learners model up to a certain point without receiving feedback from the instructor, they may continue inefficient learning from exercises.
Therefore, it can be said that it is also important to end the exercise at an appropriate time.

Common points in both groups include a large increase in $CDS$ between 10 and 20 minutes, a small increase between 20 and 40 minutes, and a large increase between 40 and 50 minutes.
This suggests that there is a cycle in the transition of similarity in this modeling exercise.
This cycle has a period of about 30 minutes, with the first cycle being between 0 and 30 minutes, during which elements such as classes, attributes, and relationships are created, and the second cycle being between 30 and 60 minutes, during which the created elements are deleted, moved, renamed, \etc
Assuming this cycle exists, the effect of this method appears from the second cycle, which involves correcting the class diagram.

Ichinohe \etal\cite{ichinohe}, who used the same \textit{KIfU} as in this paper, provide feedback on the progress during the modeling exercise using numerical values.
However, since there was a discrepancy between the learners' perception and the progress of modeling, there were also subjects whose similarity improvement was unstable.
In contrast, the similarity of the experimental group in this paper was steadily improved.

\section{Threats to Validity}
The following two points can be as potential factors that may affect the validity of the results of this research.

The first point is related to similarity.
In this research, we are using and evaluating a method that uses the similarity defined by Tanaka \etal\cite{similarity}.
Therefore, if metrics defined in other studies are used, the results may differ.

The second point is related to the complexity of the correct example used in the experiment.
In this experiment, no subjects were found who became unable to understand their own work by automatically converting the layout.
However, if the correct example becomes more complex, there is a possibility that the student's work may become incomprehensible due to the automatic layout conversion.
For this reason, it is necessary to conduct evaluations using cases with various other class diagrams in the future.

\section{Conclusion and Future Work}

Modeling exercises for software development targets are difficult for learners, and the class diagrams they create often contain many errors and differ significantly from the correct examples.
To solve this problem, this paper proposed a method to automatically convert the layout of a learner's class diagram to the layout of a class diagram created by an instructor, and implemented the proposed method as a tool.

In the experiment, an exercise was conducted, and the similarity of the experimental group that could use the automatic layout conversion function was compared with that of the control group that could not.
As a result, the similarity was higher in the experimental group than in the control group, indicating that the proposed method is effective feedback for learners.
Other noteworthy points include that the experimental group had a larger variance, and there were common points in the changes in similarity of both groups during the exercise.
The reason why the variance of similarity was larger in the experimental group is thought to be due to the mixture of subjects who needed feedback at different times.
In addition, from the changes in similarity during the exercise, a suggestion was suggested that there is a certain cycle in class diagram modeling.

Future possible research includes investigating the relationship between the feeling of understanding the reason for relationship creation and the learning effect, and conducting evaluations with increased feedback opportunities to reduce the variation in support effectiveness among learners.

\section*{Acknowledgements}
A part of this research was supported by JSPS KAKENHI Grant Number 24K15214.


\balance
\bibliographystyle{jssst}
\bibliography{sample}

\onecolumn
\appendices
\renewcommand\thesubsection{\thesection.\Roman{subsection}}
\renewcommand\thesubsectiondis{\thesection.\arabic{subsection}}

\section{Problem Used in the Experiment}
\subsection{Problem Statement (Requirements)}
Wakaba Trading Company has decided to digitize (online) product orders that were previously placed via fax and telephone.
In this system, customers will place orders directly from the homepage.
Although orders will be online, they want to continue using the three systems that have been used so far: the product management system that centrally manages the products handled, the inventory management system that manages the number of items in stock, and the customer management system that registers customer information.

Browse products should be freely available even without membership registration.
Since Wakaba Trading Company handles a wide variety of product categories and a large number of products, they want to include a product search function on the initial screen.
The search will be performed by the product management system, and when the search results are displayed, the stock information obtained from the inventory management system will also be displayed along with the product code, product name, unit price, and product category.

Customers will select the products they want to order from the searched products, enter the quantity, and add them to the cart. If the product they want to order is not found in the search results, they can return to the initial screen and search again.
At this time, an order item will be created for each ordered product within the cart.
Also, they want to display the contents of the cart on the screen every time a product is added to the cart so that the customer can check it each time.

Customers can automatically add and cancel products until the order is confirmed.
They want to ensure that the information is always confirmed whenever the contents of the cart are changed.
Once the products to be ordered are confirmed, the order details and the total amount will be displayed on the screen, and the customer will be required to enter their name, name in katakana, postal code, address, and phone number.
They want to make customer information registration mandatory for product orders.
New customers will need to enter all information, but for customers who have used this system before, a unique customer code will be assigned, and by entering that number, they will be able to place an order without re-entering the customer information they registered previously.
Once the customer information is registered, the information necessary for the order (order details and customer information) will be confirmed, and if there are no errors, the order will be registered.

When an order is registered, they want the number of ordered items to be subtracted from the stock quantity registered in the inventory management system so that the stock quantity is updated.
Once all processes are completed, they want to notify the order code, order acceptance date, and the fact that the order has been accepted, and all order processing operations will be completed.

\subsection{Instructor's Correct Example (Class Diagram)}
\begin{figure*}[h]
  \begin{center}
    \includegraphics[width=0.95\textwidth]{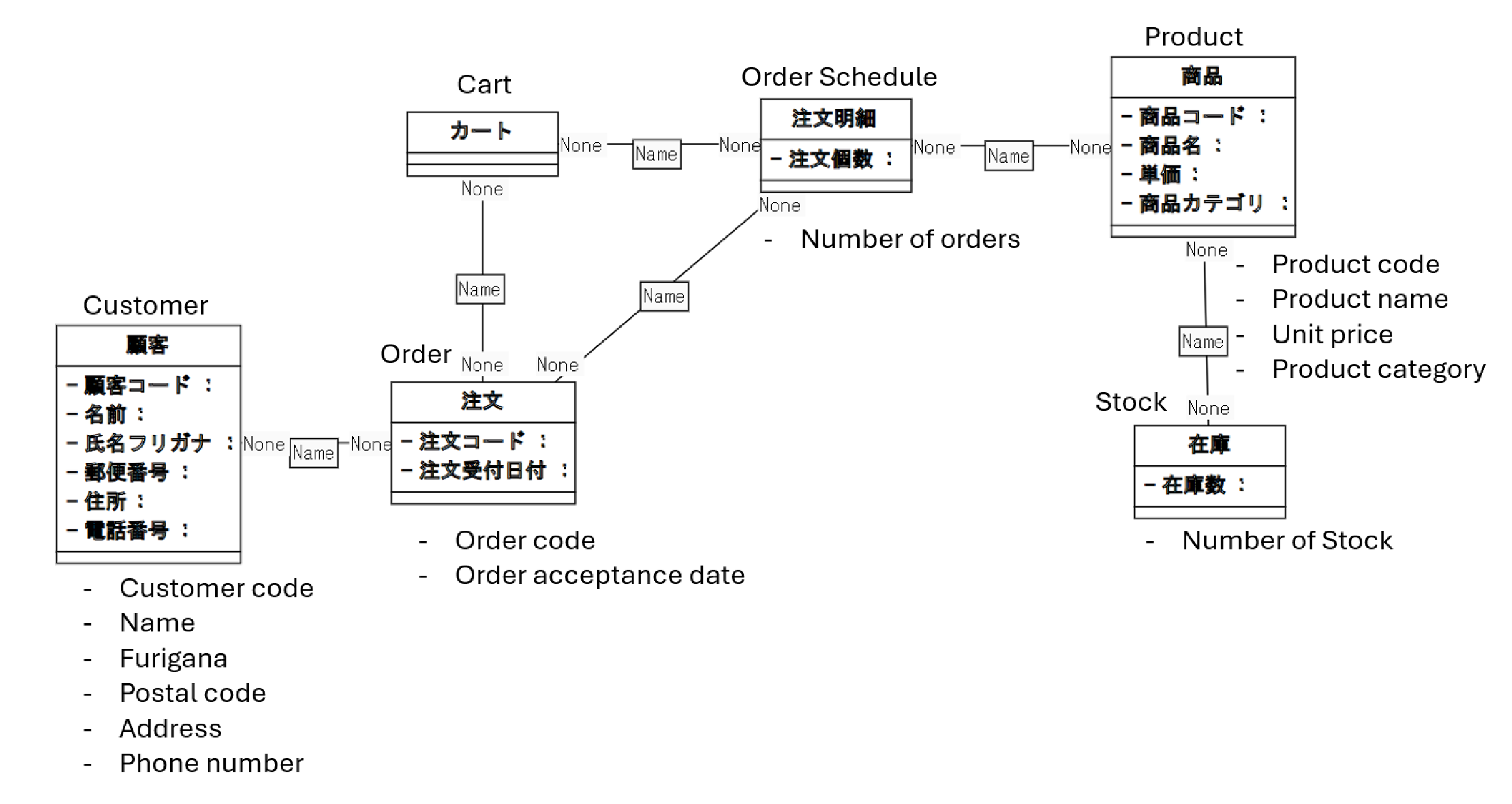}
    \label{answer}
  \end{center}
\end{figure*}

\end{document}